\documentclass[twocolumn]{IEEEtran}

\usepackage{amsmath,epsfig,amssymb,verbatim,amsopn,cite,multirow}
\usepackage{amsthm}
\usepackage{balance}
\usepackage{multirow}
\usepackage[usenames,dvipsnames]{color}
\usepackage[all]{xy}  
\usepackage{url}
\usepackage{amsfonts}
\usepackage{amssymb}
\usepackage{epsfig}
\usepackage{epstopdf}
\usepackage{bm}
\usepackage{balance}
\usepackage{graphicx}
\usepackage{subcaption}
\usepackage{footnote}
\usepackage{algorithm,algorithmic}

\usepackage[margin=15.7mm,top=18mm,bottom=25.7mm]{geometry}

\usepackage[nodisplayskipstretch]{setspace}
\newtheorem{Theorem}{Theorem}

  {\proof}{\proofend}

\newcommand{\qa}{{\bf a}}

\newcommand{\qg}{{\bf g}}

\newcommand{\qr}{{\bf r}}

\newcommand{\qw}{{\bf w}}
\newcommand{\qx}{{\bf x}}
\newcommand{\qy}{{\bf y}}

\newcommand{\qF}{{\bf F}}

\DeclareMathOperator*{\argmax}{arg\,max}

\newcommand{\SINR}{\mathsf{SINR}}

\newcommand{\Ntx}{N}
\newcommand{\Nrx}{N}

\newcommand{\dl}{\mathtt{J}}
\newcommand{\ul}{\mathtt{O}}
\newcommand{\UT}{\mathtt{UT}}
\newcommand{\UR}{\mathtt{UR}}

\newcommand{\hgmkpd}{\hat{\qg}_{mk'}^{\dl}}
\newcommand{\tgmkd}{\tilde{\qg}_{mk}^{\dl}}
\newcommand{\tgmlu}{\tilde{\qg}_{mk}^{\ul}}

\newcommand{\gmkd}{\qg_{mk}^{\dl}}

\newcommand{\hgmkd}{\hat{\qg}_{mk}^{\dl}}

\newcommand{\hgnkd}{\hat{\qg}_{nk}^{\dl}}

\newcommand{\hgmlu}{\hat{\qg}_{mk}^{\ul}}

\newcommand{\gmlu}{\qg_{mk}^{\ul}}

\newcommand{\gamdmk}{\gamma_{mk}^{\dl}}

\newcommand{\gamdmkp}{\gamma_{mk'}^{\dl}}
\newcommand{\gamuml}{\gamma_{mk}^{\ul}}

\newcommand{\betamkd}{\beta_{mk}^{\dl}}

\newcommand{\betakkdu}{\beta_{kk}^{\mathtt{U}}}

\newcommand{\betakldu}{\beta_{\ell k}^{\mathtt{U}}}

\newcommand{\betamlu}{\beta_{mk}^{\ul}}

\DeclareMathOperator{\F}{\mathbf{F}} 

\DeclareMathOperator{\aaa}{\mathbf{a}}

\DeclareMathOperator{\K}{\mathcal{K}}

\DeclareMathOperator{\MM}{\mathcal{M}}

\DeclareMathOperator{\THeta}{\boldsymbol{\theta}}

\title{\fontsize{0.83cm}{1cm}\selectfont   Cell-Free Massive MIMO Surveillance Systems}
\author{Zahra Mobini$^\dag$, Hien Quoc Ngo$^\dag$, Michail Matthaiou$^\dag$, and Lajos Hanzo$^\ddag$\\
\small{
$^\dag$Centre for Wireless Innovation (CWI), Queen's University Belfast, U.K.\\
 $^\ddag$Department of Electronics and Computer
Science, University of Southampton, SO17 1BJ Southampton, U.K.\\
Email:\{zahra.mobini, hien.ngo, m.matthaiou\}@qub.ac.uk}, lh@ecs.soton.ac.uk}\normalsize
\allowdisplaybreaks
\allowdisplaybreaks
\begin{document}

\bstctlcite{IEEEexample:BSTcontrol}
\maketitle

\begin{abstract}
Wireless surveillance, in which untrusted communications links are proactively monitored by legitimate agencies,
has started to garner a lot of interest for enhancing the national
security. In this paper, we propose a new cell-free massive multiple-input multiple-output (CF-mMIMO) wireless surveillance system,
where a large number of distributed multi-antenna aided legitimate
monitoring nodes (MNs) embark on either observing or jamming untrusted communication links.
To facilitate concurrent observing and jamming, a subset of
the MNs is selected for monitoring the untrusted transmitters
(UTs), while the remaining MNs are selected for jamming
the untrusted receivers (URs). We analyze the
performance of CF-mMIMO wireless surveillance and derive a
closed-form expression for the monitoring success probability of MNs.
We then propose a greedy algorithm for the observing vs, jamming mode assignment of MNs,
followed by the conception of a jamming transmit power allocation algorithm for maximizing
the minimum monitoring success probability concerning all the
UT and UR pairs based on the associated long-term channel state information
 knowledge. In conclusion, our proposed CF-mMIMO system is capable of 
significantly improving the performance of the MNs
compared to that of the state-of-the-art baseline. In  scenarios of a
mediocre number of MNs, our proposed scheme provides an 11-fold improvement in the minimum monitoring success probability compared to its co-located mMIMO  benchmarker.

\let\thefootnote\relax\footnotetext{This work is a contribution by Project REASON, a UK Government funded project under the Future Open Networks Research Challenge (FONRC) sponsored by the Department of Science Innovation and Technology (DSIT). The work of H. Q. Ngo was supported by the U.K. Research and Innovation Future Leaders Fellowships under Grant MR/X010635/1. The work of M. Matthaiou was
supported by the European Research Council
(ERC) under the European Union’s Horizon 2020 research
and innovation programme (grant agreement No. 101001331).}
\end{abstract}
\vspace{-1.3em}
\section{Introduction}
While the development of wireless communication systems has been dramatically improved throughout  the wireless generations, there is an increased demand for further improved wireless security.
More specifically, infrastructure-free or user-controlled wireless communication networks, such as   device-to-device (D2D)  and mobile ad-hoc communications, make public safety more vulnerable to threats. Unauthorised or malicious users may misuse the available wireless infrastructure-free networks to perform illegal activities, commit cyber crime, and jeopardize public safety. This has motivated researchers to propose new surveillance methods, termed as proactive monitoring, to allow the legitimate monitor  to observe or even degrade  the quality of untrusted communication~\cite{Zhang:MAg:2018}.

In proactive monitoring, the legitimate monitor operates in a full-duplex (FD) mode and sends a jamming signal to interfere with the reception of the UR, thereby  degrading the rate of the untrusted link. This improves the   monitoring performance~\cite{Zhong:TWC:2017}. The applications and different features of proactive monitoring  in wireless information surveillance have been  studied in the literature in various scenarios, such as  MIMO systems~\cite{Zhong:TWC:2017,Feizi:TCOM:2020}, unmanned aerial vehicle (UAV) networks~\cite{Mobini:ICC:2018}, cognitive radio networks~\cite{Ge:ITJ:2022},  relay-aided monitoring links~\cite{Moon:2018:TWC,Li:2018:GLOBECOM}, and non-orthogonal multiple access (NOMA) networks~\cite{Xu:TVT:2020}. However, current studies tend to investigate the simple scenario of a single untrusted link and/or a single  legitimate monitor for direct monitoring.  
These  assumptions are optimistic, because realistic systems are likely to have more than one untrusted communication links in practice.
But in the face of a distributed deployment of untrusted pairs, it is impractical to arrange for the direct monitoring of each and every untrusted pair by relying on a single monitor. Xu and  Zhu~\cite{Xu:TWC:2022}  studied proactive monitoring  using a single monitor observing multiple untrusted pairs in  scenarios associated with quality-of-service-constrained untrusted links, while Li \emph{et al.} \cite{Li:LSP:2018} used    proactive monitoring with relaying features to increase the signal-to-interference-plus-noise ratio 
 (SINR) of the untrusted link,  which results in higher rate of the untrusted link, and hence higher observation rate.
Moreover, Moon \emph{et al.}~\cite{Moon:TWC:2019} looked into  proactive monitoring, relying on a group of single-antenna intermediate relay nodes harnessed  for supporting a distant legitimate monitor, which act either as a jammer or as an observer node.  However, Moon \emph{et al.}~\cite{Moon:TWC:2019} only focused on the single-untrusted-link scenario  under  the  overly optimistic assumption of having perfect global channel state information (CSI) knowledge of all links at the  monitor. Therefore, the study of how to efficiently perform a realistic surveillance operation using multiple monitors in the presence of multiple untrusted pairs is meaningful and important, but is still an open problem at the time of writing.

To address  the need for reliable information surveillance in complex  practical scenarios, we are inspired by the emerging concept of CF-mMIMO to  propose a new proactive monitoring system, termed as \emph{CF-mMIMO surveillance}. CF-mMIMO breaks the concept of cells or cell boundaries by using numerous distributed access points  for simultaneously serving a  smaller number of users~\cite{Hien:cellfree}. It
 enjoys all benefits granted by the mMIMO and coordinated multi-point concepts relying on joint transmission, and hence, can guarantee ubiquitous coverage for all the users.
Our CF-mMIMO information surveillance system  compromises a large number of spatially distributed legitimate MNs, which coherently perform surveillance  of multiple untrusted pairs distributed over a geographically wide area.

In our system, for any given untrusted pair, there are typically several MNs in close proximity, and hence, high macro diversity gain can be achieved. As a consequence,   CF-mMIMO surveillance  is expected to offer an improved monitoring performance compared to its single-monitor (co-located) mMIMO based counterpart.
This is achieved by CF-mMIMO, because a virtual FD mode can be realized by relying on half-duplex MNs. More precisely, in the  system, two types of MNs are relied upon: 1) a specifically selected subset of the MNs is purely used for observing the UTs; 2) the rest of the MNs cooperatively jam the URs. Since the MNs are distributed across the area, the node-to-node interference in our proposed system is significantly reduced compared to a conventional FD monitoring/jamming system.
The observing mode or jamming mode and MN transmit power  can be dynamically adjusted for maximizing the overall monitoring performance based on only long-term CSI.

The main technical contributions and key novelty of this paper are summarized
as follows:
\begin{itemize}
    \item We propose a novel wireless surveillance system, which is based on the CF-mMIMO technology and  either observing or jamming mode assignment. In particular, by assuming realistic imperfect CSI knowledge, we characterize  the  monitoring success probability  of CF-mMIMO surveillance system  over multiple untrusted pairs. 
    \item  We formulate an optimization problem for maximizing the minimum monitoring success probability  of all the untrusted pairs subject to a realistic per-MN average transmit power constraint. The proposed  power control solution can be found by solving a sequence of linear programs. We also propose a greedy algorithm for MN mode assignment. 
    \item Numerical results show that the proposed CF-mMIMO surveillance system substantially improves the monitoring performance over that of   mMIMO based proactive monitoring systems relying on FD operation, where  all MNs are co-located as an antenna array, which simultaneously perform observation and jamming.
\end{itemize}

\textit{Notation:} We use bold upper case letters to denote matrices, and lower case letters to denote vectors.  The superscripts $(\cdot)^*$, $(\cdot)^T$ and $(\cdot)^\dag$ stand for the conjugate, transpose, and conjugate-transpose (Hermitian), respectively. The zero mean circular symmetric complex Gaussian distribution having variance $\sigma^2$ is denoted by $\mathcal{CN}(0,\sigma^2)$, while $\mathbf{I}_N$ denotes the $N\times N$ identity matrix.  Finally, $\mathbb{E}\{\cdot\}$ denotes the statistical expectation.

\vspace{-0.8em}
\section{System model}~\label{sec:Sysmodel}
We consider the surveillance scenario of multiple untrusted communication links, including  $M$ MNs and $K$ untrusted communication pairs. Each  UT and UR is equipped with a single antenna, while each MN is equipped with $N$  antennas.  All MNs, UTs,  and URs are half-duplex devices.  We assume that all MNs are connected to the central processing unit (CPU) via  fronthaul links.  The MNs can switch between observing mode, where they receive untrusted messages, and jamming mode, where they send jamming signals to disrupt URs. The assignment of each mode to its corresponding MN is designed to maximize the monitoring success probability (see Section~\ref{sec:SE}). The binary variable representing the mode assignment for each MN $m$ is expressed as
$a_{m}= 1$ ($a_{m}= 0$) if MN $m$ operates in the jamming mode (observing mode).

Denote   the jamming channel (observing channel)  between the $m$-th MN and the $k$-th UR ($k$-th UT) by  $\gmkd\in\mathbb{C}^{\Ntx \times 1}$  ($\gmlu\in\mathbb{C}^{\Nrx \times 1}$), $\forall k  \in  \K\triangleq \{1,\dots,K\}, m \in \MM \triangleq \{1, \dots, M\}$, respectively.
We model it by $\gmkd=\sqrt{\betamkd}\tgmkd,~(\gmlu=\sqrt{\betamlu}\tgmlu)$, where $\betamkd$ ($\betamlu$) is the large-scale fading coefficient and $\tgmkd\in\mathbb{C}^{\Ntx \times 1}$ ($\tgmlu\in\mathbb{C}^{\Ntx \times 1}$) is the small-scale fading vector associated with independent and identically distributed (i.i.d.) $\mathcal{CN} (0, 1)$ random variables (RVs). Also, the channel gain between the $l$-th UT  and the $k$-th UR  is  $h_{\ell k}=(\betakldu)^{1/2}\breve{h}_{\ell k}$, where $\betakldu$ is the large-scale fading coefficient and $\breve{h}_{\ell k}$ represents small-scale fading, distributed as $\mathcal{CN}(0,1)$. We note that $h_{kk}$ represents the channel coefficient of the $k$-th
untrusted link spanning from the $k$-th UT to the
$k$-th UR, $ \forall k \in \K$.
Finally, the channel matrix between MN $m$ and MN $i$, $\forall m,i\in\MM$, is denoted by $\qF_{mi}\in \mathbb{C}^{\Nrx\times\Ntx}$. We assume that the elements of $\qF_{mi}$, for $i\neq m$, are  i.i.d. $\mathcal{CN}(0,\beta_{mi})$ RVs and $\F_{mm} = \bold{0}, \forall m$.
Note that the channels $\gmkd$ and $\gmlu$
may be estimated at the legitimate MN by overhearing the
pilot signals sent by UR $k$ and UT $k$, respectively~\cite{Moon:2018:TWC}.
We denote   the estimates of $\gmkd$  and $\gmlu$ by $\hgmkd$ and $\hgmlu$, respectively where $\hgmkd \sim \mathcal{CN}(\bold{0},\gamdmk \mathbf{I}_N)$, and $\hgmlu \sim \mathcal{CN}(\bold{0},\gamuml \mathbf{I}_N)$. Since it is difficult  for the legitimate MNs to obtain the CSI of untrusted links, we assume that  $h_{k\ell}$ is unknown to the  MNs.

All the UTs simultaneously send independent
untrusted messages to their corresponding receivers over the
same frequency band. The signal  transmitted from  UT $k$ is denoted by $x_{k}^\UT  = \sqrt{\rho_\UT} s_{k}^{\UT}$, 
where $s_{k}^\UT$, with $\mathbb{E}\left\{|s_{k}^\UT|^2\right\}=1$, and $\rho_\UT$ represent  the transmitted symbol and  the  normalized transmit power at each UT, respectively.
In addition, at the same time, the MNs in jamming mode intentionally send jamming signals to interfere with the untrusted transmission links. This reduces the achievable data rate at the URs, while improving the monitoring success probability. In particular, the MNs in  jamming mode use the maximum-ratio (MR)  transmission technique (a.k.a. conjugate beamforming) to jam the reception of the URs. 
Denote the jamming symbol intended for the untrusted link $k$  by $s_k^\dl$, which is a RV with zero mean and unit variance. With MR precoding, the $\Ntx\times 1$  signal vector transmitted by MN $m$  can be expressed as $\qx_{m}^{\dl}
= a_m\sqrt{\rho_\dl}\sum_{k \in \mathcal{K}} \sqrt{\theta_{mk}} \left(\hgmkd\right)^*
s_{k}^{\dl}$,
where  $\rho_\dl$ is the maximum normalized transmit power at each MN in the jamming mode. Moreover, $ \theta_{mk}$ is the  power control coefficient chosen to satisfy the power constraint $\mathbb{E}\left\{\|\qx_{m}^{\dl}\|^2\right\} \leq \rho_\dl$ at each MN, which can be further written as
\vspace{-0.2em}
\begin{align}
\label{DL:power:cons}
a_m\sum_{k\in\mathcal{K}} \gamdmk \theta_{mk} \leq \frac{1}{\Ntx}, \forall m.
\end{align}
Accordingly, the signal received  by UR $k$ can be written as
\vspace{-0.1em}
\begin{align}~\label{eq:ykdl}
&y_k^{\UR}
= h_{kk}x_{k}^{\UT}+ \sum_{\ell\in \mathcal{K}, \ell \neq k}h_{\ell k}x_{\ell}^{\UT}
+\!\sqrt{\rho_\dl}\times
\nonumber\\
&
\sum_{m \in \mathcal{M}}
\sum_{k'\in\mathcal{K}} a_m\sqrt{\theta_{mk'}}
\left(\gmkd\right)^T\left(\hgmkpd\right)^*
s_{k'}^{\dl}+w_{k}^{\UR},
\end{align}
where $w_{k}^{\UR}\sim\mathcal{CN}(0,1)$ is the additive white Gaussian noise (AWGN) at UR $k$. We notice that the second term in~\eqref{eq:ykdl}, is the interference caused by other UTs due to their concurrent transmissions over the same frequency band while the third term is the interference emanating from the MNs in the jamming mode.

The MNs in the observing mode, i.e., MNs  with $a_m=0, \forall m$, receive the transmit signals from all UTs. The received signal $\qy_{m}^{\ul}\in\mathbb{C}^{\Nrx \times 1}$ at MN $m$ in the observing mode is given by
\vspace{-0.1em}
\begin{align}\label{eq:ymul}
&\qy_{m}^{\ul}
=
\sqrt{\rho_\UT}\sum_{k\in \mathcal{K}}(1-a_m)\qg_{mk}^{\ul} s_{k}^{\UT}
+\sqrt{\rho_\dl}\times\\
&
~\sum_{i\in\mathcal{M}}\sum_{k\in \mathcal{K}}
(1-a_m)a_i \sqrt{\theta_{ik}}
\qF_{mi}
(\hat{\qg}_{ik}^\dl)^*s_k^\dl
+(1-a_m)\qw_{m}^{\ul},\nonumber
\end{align}
where $\qw_{m}^{\ul}$ is the $\mathcal{CN}(\bold{0}, \mathbf{I}_N)$ AWGN vector. We note  from~\eqref{eq:ymul}  that if MN $m$ does not operate in the observing mode, i.e., $a_m=1$, it does not receive any signal, i.e., $\qy_{m}^{\ul}=\boldsymbol{0}$. 

Then, MN $m$ in the observing mode performs MR combining (MRC) reception  by partially equalizing the received signal in~\eqref{eq:ymul} based on the  Hermitian of the local channel estimates as $(\hgmlu)^\dag\qy_{m}^{\ul}$. The resultant signal is then forwarded to the CPU for detecting the untrusted signals, where the receive combiner sums up the equalized  signals. The aggregated received signal for UT $k, \forall k,$ at  the CPU can be written as
\vspace{-0.2em}
\begin{align}\label{eq:rul}
	r_{k}^{\ul}=\sum_{m=1}^{M} (\hgmlu)^\dag\qy_{m}^{\ul}. 
\end{align}
Finally, the observed information $s_{k}^{\UT}$ can be detected from $\qr_{k}^{\ul}$. 

\subsubsection{Effective SINR of the Untrusted Communication Links}
We define the effective noise as
\begin{align}~\label{eq:ykdl3}
&\tilde w_{k}^{\UR}
=   \sqrt{\rho_\UT} \sum_{\ell\in \mathcal{K}, \ell \neq k}h_{\ell k} s_{\ell}^{\UT}
\nonumber\\
&\hspace{0em}
+\sqrt{\rho_\dl} \sum_{m \in \mathcal{M}}\!
\sum_{k'\in\mathcal{K}}\!\! a_m\sqrt{\theta_{mk'}}
\left(\gmkd\right)^T\left(\hgmkpd\right)^*
s_{k'}^{\dl}\!+w_{k}^{\UR},
\end{align}
and reformulate the  signal received at UR $k$ in~\eqref{eq:ykdl}  as
 \begin{align}~\label{eq:ykdl2}
 y_k^{\UR}
 = \sqrt{\rho_\UT} h_{kk}s_{k}^{\UT}+\tilde w_{k}^{\UR}.
 \end{align}
 Since $s_{\ell}^{\UT}$ is independent of $s_{k}^{\UT}$ for any $\ell \neq k$,  the first term of the effective noise~\eqref{eq:ykdl3} is uncorrelated with the first term in~\eqref{eq:ykdl2}. Moreover, the second and third terms of~\eqref{eq:ykdl3} are uncorrelated with the first term of~\eqref{eq:ykdl2}. Therefore, the effective noise $\tilde w_{k}^{\UR}$ and the input RV $x_{k}^{\UT}$  are uncorrelated. Accordingly, we obtain the closed-form expression for the effective  SINR of the untrusted link $k$ as in the following theorem.
\begin{Theorem}\label{Theorem:SE:SUS}
The effective SINR  of the untrusted link $k$ can be formulated as 
\vspace{-0.5em}
\begin{align}\label{eq:SINRSR}
    &\SINR_{k}^\UR  (\qa, \boldsymbol \theta) = \frac{ \rho_{\UT}|  h_{kk}|^2}{\xi_k(\qa,\boldsymbol{\theta})},
\end{align}
where
\vspace{-0.2em}
\begin{align}\label{eq:zeta}
\xi_k(\qa,\boldsymbol{\theta})
=& \rho_{\UT}\!\!\!\sum_{\ell\in \mathcal{K}\setminus k} \!\betakldu\!+{\rho_\dl}N
\sum_{k'\in\mathcal{K}}\sum_{m \in \mathcal{M}}
 a_m\theta_{mk'}\betamkd\gamdmkp  
\nonumber\\
 &\quad+{\rho_\dl}N^2\Big(\sum_{m \in \mathcal{M}}a_m \sqrt{\theta_{mk}} \gamdmk\! \Big)^2+1,
 \end{align}
with
$\qa \triangleq \{a_m\}$ and $\boldsymbol{\theta}\triangleq \{\theta_{mk}\}$, $\forall m,k$,  respectively.
\end{Theorem}
\textit{Proof:} See Appendix~\ref{ProofTheorem:SE:SUS}.$\hspace{14em}\blacksquare$
\subsubsection{Effective SINR for Observing}
The CPU detects the observed information $s_{k}^{\UT}$ from $	\qr_{k}^{\ul}$ in~\eqref{eq:rul}. We assume that it does not have instantaneous CSI knowledge of the observing, jamming, and untrusted channels and  uses only statistical knowledge when performs detection. By employing  the use-and-then-forget capacity-bounding methodology~\cite{Hien:cellfree}, the received SINR   for observing the untrusted link $k$ can be obtained as in the following theorem. 
\begin{Theorem}\label{Theorem:SE:CPU}
The received SINR for the $k$-th untrusted link at the CPU is given by
\vspace{-0.0em}
\begin{align}\label{eq:SINRMA}
&\SINR_{k}^\ul(\qa,\boldsymbol{\theta})=
\nonumber
\\
&\frac{
	\Nrx \rho_{\UT} \Big(\sum\limits_{\substack{m\in\mathcal{M}}} (1-a_m )  \gamuml \Big)^2
	}
	{\mu_k+
		\rho_\dl\Ntx
		\!\sum\limits_{\substack{m\in\mathcal{M}}}
		\sum\limits_{\substack{i\in\mathcal{M}}}
		\sum\limits_{\ell\in\mathcal{K}}
		\!
		(1-a_m) a_{i}\theta_{i\ell}   \gamuml \beta_{mi} \gamma_{i\ell}^{\dl}
},
\end{align}
with
\vspace{-0.4em}
\begin{align*}
\mu_k\triangleq&\rho_{\UT}
		\! \sum\limits_{\substack{m\in\mathcal{M}}}
		 \sum\limits_{\ell\in\mathcal{K}}\!
				(1-a_m) 
	\beta_{m\ell}^{\ul}
		\gamma_{mk}^{\ul}
	+\sum\limits_{\substack{m\in\mathcal{M}}}
		(1\!\!-\!a_m) \gamuml.
  \end{align*}

\end{Theorem}
\textit{Proof:} See Appendix~\ref{ProofTheorem:SE:CPU}. $\hspace{14em}\blacksquare$


\subsubsection{Monitoring Success Probability}
To achieve  reliable detection at UR $k$, UT $k$ varies its transmission rate according to $\SINR_{k}^\UR$. Hence, if $\SINR_{k}^\ul\geq\SINR_{k}^\UR$,  the CPU can also reliably detect the information of the untrusted link $k$. On the other hand, if $\SINR_{k}^\ul\leq\SINR_{k}^\UR$, the CPU may detect this information at a high probability of error. Therefore, the following indicator function can be designed for characterising the event of
successful monitoring at the CPU~\cite{Zhong:TWC:2017}
\begin{align*}
X_k = \begin{cases} \displaystyle 1 & \text {if } \SINR_{k}^\ul\geq\SINR_{k}^\UR,\\ \displaystyle 0 & \text {otherwise},
\end{cases}
\end{align*}
where $X_k=1$ and $X_k=0$ indicate monitoring success and failure events for the untrusted link $k$, respectively.
Thus, a suitable performance metric for  monitoring
each untrusted communication link $k$ is the \textit{monitoring success probability} $\mathbb {E}\{X_k\}$, defined as 
\begin{equation} \label{eq:non-outage}
\mathbb {E}\{X_k\}= \text {Pr}\left (\SINR_{k}^\ul\geq\SINR_{k}^\UR\right)\!.
\end{equation}
From~\eqref{eq:SINRSR},~\eqref{eq:SINRMA}, and~\eqref{eq:non-outage} we have
\vspace{-0.2em}
\begin{equation}
\mathbb {E}\{X_k\}=\mathbb {P}\Big(| h_{kk}|^2\leq \frac{\SINR_{k}^\ul \xi_k}{\rho_{\UT}}\Big).
\end{equation}
Using the cumulative distribution function (CDF) of the exponentially distributed RV  $|h_{kk}|^2$, the monitoring success probability  of our CF-mMIMO surveillance system can be expressed in closed form as
\vspace{-0.2em}
\begin{equation} \label{eq:Nonout}
\mathbb {E}\{X_k\}=1-\exp \left(-\frac{\SINR_{k}^\ul \xi_k}{\betakkdu\rho_{\UT}}\right). \end{equation}
\vspace{-1.2em}
\section{Max Min  Optimization}
\label{sec:SE}
In this section, we aim for maximizing the lowest probability of successful monitoring by  optimizing the observing and jamming mode assignment vector $\aaa$ and power control coefficient vector $\THeta$ under the constraint of the transmit power at each MN in~\eqref{DL:power:cons}. More precisely,  we formulate an optimization problem as
\vspace{-0.7em}
\begin{subequations}\label{P:SE2}
	\begin{align}
		\underset{\{ \qa,\boldsymbol \theta\}}{\mathrm{max}}\,\, &\hspace{1em}
		\underset{k\in\mathcal{K}} \min \,\,\mathbb {E}\{X_k ({\boldsymbol{\qa,\theta}}) \}
		\\
		\mathrm{s.t.} \,\,
		& \hspace{1em} a_m\sum_{k\in\mathcal{K}} \gamdmk \theta_{mk} \leq \frac{1}{\Ntx},~~  m\in\mathcal{M},\label{opt:cons1}\\
  &\hspace{1em}\theta_{mk}\geq 0, ~~ m\in\mathcal{M},~k\in\mathcal{K}, \label{opt:cons2}\\
  &\hspace{1em}a_{m} \in \{0,1\}, ~~ m\in\mathcal{M}. 
   \end{align}
\end{subequations}
Since  $1-\exp(-x)$ is a monotonically increasing function of $x$, the problem~\eqref{P:SE2} is equivalent to the following problem
\begin{subequations}\label{P:SE3}
\	\begin{align}
		&\underset{\{\qa,\boldsymbol \theta\}}{\mathrm{max}}\,\, \hspace{1em}
		\underset{k\in\mathcal{K}} \min \,\,\SINR_{k}^\ul(\qa,\boldsymbol \theta)\xi_k(\qa,\boldsymbol \theta)
		\\
  &~\text {s.t.} \hspace{2em}~\eqref{opt:cons1},~\eqref{opt:cons2}.
	\end{align}
 \end{subequations}
 %
Problem~\eqref{P:SE3} is  a challenging combinatorial problem. Therefore, for MN mode assignment, we only focus on  a heuristic greedy method which simplifies the computation, while providing a significant successful monitoring performance gain.
\vspace{-0.4em}
\subsection{Greedy MN Mode Assignment for  Fixed Power Control}
Let $\mathcal{A}_{\ul}$ and $\mathcal{A}_{\dl}$ denote the sets containing the indices of MNs in observing mode, i.e., MNs with $a_m=0$, and MNs in jamming mode, i.e., MNs with $a_m=1$, respectively. 
In addition, $\mathbb {E}\{X_k({A}_{\ul}, \mathcal{A}_{\dl})\}$ presents the dependence of the monitoring success probability  on the different choices of MN mode assignments. Our greedy algorithm of MN mode assignment is shown in \textbf{Algorithm~\ref{alg:Grreedy}}. All MNs are
initially assigned to observing mode, i.e., $\mathcal{A}_{\ul}=\mathcal{M}$ and $\mathcal{A}_{\dl}=\emptyset$. Then, in each iteration,  one MN switches into jamming mode for maximizing the minimum monitoring success probability~\eqref{eq:Nonout} among the untrusted links, until there is no more improvement.

\vspace{-0.6em}
\subsection {Power Control}
For a given MN mode assignment, the optimization problem~\eqref{P:SE3}  reduces to the
 power control problem. Using~\eqref{eq:zeta},~\eqref{eq:SINRMA}, and~\eqref{P:SE3} the max-min fairness
problem is now formulated as
\vspace{0.2em}
\begin{subequations}\label{eq:PAoptimization1}
\begin{align}
&\max _{\boldsymbol{\theta}}~\min \limits _{\forall k\in\mathcal{K}}  \frac {\xi_k(\boldsymbol{\theta})}
{	\!\mu_k\!+\!\rho_\dl\Ntx\!\!\!
		\!\sum\limits_{\substack{m\in\mathcal{M}}}
		\!\sum\limits_{\substack{i\in\mathcal{M}}}
		\sum\limits_{\ell\in\mathcal{K}}
		\!
		(\!1\!-\!a_m\!) a_{i}\theta_{i\ell}   \gamuml \beta_{mi} \gamma_{i\ell}^{\dl}
		} \\[0.5pt]
&\text {s.t.} \hspace{2em}~\eqref{opt:cons1},~\eqref{opt:cons2}.
\end{align}
\end{subequations}
By introducing the slack variable $\varrho$, we  reformulate~\eqref{eq:PAoptimization1} as
\vspace{-0.1em}
\begin{subequations}~\label{eq:PAoptimization}
\begin{align} 
&\max _{\{\boldsymbol{\theta}, \varrho\}}~ \varrho \\
&\text {s.t.} \hspace{2em}	\rho_\dl\Ntx
		\!\sum\limits_{\substack{m\in\mathcal{M}}}
		\sum\limits_{\substack{i\in\mathcal{M}}}
		\sum\limits_{\ell\in\mathcal{K}}
		(1-a_m) a_{i}\theta_{i\ell}   \gamuml \beta_{mi} \gamma_{i\ell}^{\dl}\nonumber\\
          &\hspace{8em}- \frac{1}{\varrho}{\xi}_k(\boldsymbol{\theta}) + \mu_k\leq 0,
~~\forall k\in\mathcal{K}, \label{eq:cons_zet}
 \\
&\hspace{2em}~\eqref{opt:cons1},~\eqref{opt:cons2}.
\end{align}
\end{subequations}
To arrive at a  computationally more efficient formulation, we use the inequality $\left(\sum_{m \in \mathcal{M}}\sqrt{\theta_{mk}}\gamdmk\right)^2 \geq \sum_{m \in \mathcal{M}}\theta_{mk}(\gamdmk)^2$ and replace the constraint~\eqref{eq:cons_zet}  by
\vspace{-0.0em}
\begin{align}
\rho_\dl\Ntx
		\!\sum\limits_{\substack{m\in\mathcal{M}}}
		\sum\limits_{\substack{i\in\mathcal{M}}}
		\sum\limits_{\ell\in\mathcal{K}}
		(1-a_m) a_{i}\theta_{i\ell}   \gamuml \beta_{mi} \gamma_{i\ell}^{\dl}\nonumber\\
         - \frac{1}{\varrho}\tilde{\xi}_k(\boldsymbol{\theta}) + \mu_k\leq 0,
~~\forall k\in\mathcal{K},
\end{align}
where
\vspace{-0.1em}
\begin{align}\label{eq:zetat}
&\tilde{\xi}_k\!(\boldsymbol{\theta}) 
\!=\!\rho_{\UT}\!\!\!\sum_{\ell\in \mathcal{K}\setminus k} \!\betakldu\!+\!
{\rho_\dl}N
\!\sum_{k'\in\mathcal{K}}\sum_{m \in \mathcal{M}}\!
a_m\theta_{mk'}\betamkd\gamdmkp
\nonumber\\
&\qquad\quad+{\rho_\dl} N^2\sum_{m \in \mathcal{M}}a_m\theta_{mk}(\gamdmk)^2
+
1.
 \end{align}

Now, for a fixed $\varrho$, all inequalities involved in~\eqref{eq:PAoptimization} are linear,  hence the program~\eqref{eq:PAoptimization} is quasi–linear. 
Since the second constraint in~\eqref{eq:PAoptimization} is an
increasing functions of $\varrho$, the solution to the optimization problem is obtained by harnessing a line-search over
$\varrho$ to find the maximal feasible value.
As a consequence, we use the  bisection
method in \textbf{Algorithm 2} to obtain the solution. 
\begin{algorithm}[!t]
\caption{Greedy MN Mode Assignment}
\begin{algorithmic}[1]
\label{alg:Grreedy} 
\STATE
\textbf{Initialize}: Set  $\mathcal{A}_{\ul}=\mathcal{M}$ and $\mathcal{A}_{\dl}=\emptyset$. Set iteration index $i=0$.
\STATE Calculate $\Pi^\star[i]=  		\underset{k\in\mathcal{K}} \min \,\,\mathbb {E}\{X_k (\mathcal{A}_{\ul}, \mathcal{A}_{\dl})\}$
\REPEAT
\FORALL{$m \in \mathcal{A}_{\ul}$}
\STATE Set $\mathcal{A}_{s}=\mathcal{A}_{\ul} \setminus m$
\STATE  Calculate $\Pi_m=  		\underset{k\in\mathcal{K}} \min \,\,\mathbb {E}\{X_k (\mathcal{A}_{s}, \mathcal{A}_{\dl}\bigcup m)\}$\\
\ENDFOR
\STATE Set $\Pi^\star[i+1]= \underset{m\in\mathcal{A}_{\ul}} \max \,\,\Pi_m$\\
\STATE $e=|\Pi^\star[i+1]- \Pi^\star[i]|$ 
\IF{$e \geq e_{\min}$ }
\STATE Select MN $m^\star=\argmax_{m\in\mathcal{A}_{\ul}}\{\Pi_m\}$
\STATE {Update $\mathcal{A}_{\dl}=\{\mathcal{A}_{\dl}\bigcup m^{\star}\}$ and $\mathcal{A}_{\ul}=\mathcal{A}_{\ul}\setminus m^{\star}$}
\ENDIF
\STATE Set $i=i+1$
\UNTIL{ $e < e_{\min}$ }
\RETURN $\mathcal{A}_{\ul}$ and $\mathcal{A}_{\dl}$, i.e., the indices of MNs in observing mode and jamming mode, respectively.
\end{algorithmic}
\end{algorithm}

\begin{algorithm}[t]\label{Alg:PA}
\caption{Bisection Method for Max-Min Power Control }
\begin{algorithmic}[1]
\STATE Initialization of $\varrho_{\min}$ and $\varrho_{\max}$, where $\varrho_{\min}$ and $\varrho_{\max}$ define a range of relevant values of the objective function in~\eqref{eq:PAoptimization1}. Initial line-search accuracy  $\epsilon$.
\REPEAT
    \STATE Set $\varrho:=\frac{\varrho_{\min}+\varrho_{\max}}{2}$. Solve the following convex feasibility program
    \begin{align}\label{eq:Feasibility}
    \begin{cases} \hspace {0.0 cm}
&\!{{\mu_k\!+\!\rho_\dl\Ntx\!\!
		\!\sum\limits_{\substack{m\in\mathcal{M}}}
		\sum\limits_{\substack{i\in\mathcal{M}}}
		\sum\limits_{\ell\in\mathcal{K}}
		\!
		(\!1\!-\!a_m\!) a_{i}\theta_{i\ell}   \gamuml \beta_{mi} \gamma_{i\ell}^{\dl}}} 
  \\
  &~~-{\frac{1}{\varrho}\tilde{\xi}_k(\boldsymbol{\theta}) \leq 0
~~\forall k\in\mathcal{K}},
  \\
&a_m\sum_{k\in\mathcal{K}} \gamdmk \theta_{mk}\leq \frac{1}{\Ntx}, ~ \forall m\in\mathcal{M}, \\
&\!\theta_{mk} \geq 0, ~ \forall k\in\mathcal{K}, ~ \forall m\in\mathcal{M}.
 \end{cases} 
\end{align}
\STATE If problem~\eqref{eq:Feasibility} is feasible, then set $\varrho_{\min}:=\varrho$, else set $\varrho_{\max} :=\varrho$.
\UNTIL{ $\varrho_{\max}-\varrho_{\min}<\epsilon$ }
\end{algorithmic}
\end{algorithm}
\vspace{-1.4em}
\section{Numerical Results}

In this section, numerical results are presented for studying the performance of the proposed CF-mMIMO surveillance system. Moreover,  the performance of CF-mMIMO is  compared to that of a  co-located mMIMO surveillance system, where all the $M$ MNs are located at the centre of the coverage area as a single virtual FD large MN.

We consider a CF-mMIMO surveillance system, where the MNs, UTs, and URs are randomly distributed in an area of  $1 \times 1$ km${}^2$ having wrapped around edges to reduce the boundary effects. The distances between adjacent MNs are at least $80$ m and between each untrusted pair is within the range of $80-160$ m. Moreover, we set the number of antennas to $N=2$ and  the channel bandwidth to $B=50$ MHz. The maximum transmit power for each MN and each UT is $1$ W and $250$ mW, respectively, while the corresponding normalized maximum transmit powers ${\rho}_\dl$, ${\rho}_\UT$, can be calculated upon dividing these powers by the noise power of $-92$ dBm. The large-scale fading coefficient $\beta_{mk}$ is represented by $\beta_{mk} = 10^{{\text{PL}_{mk}^d}/{10}}10^{{A_{mk}}/{10}}$,
where the first term models the path loss, and the second term models the shadow fading with $A_{mk}\in\mathcal{N}(0,4^2)$ (in dB), respectively;  $\text{PL}_{mk}^d$ (in dB) is calculated as \cite{emil20TWC} $\text{PL}_{mk}^d = -30.5-36.7\log_{10}\left(\frac{d_{mk}}{1\,\text{m}}\right),$  
and the correlation between the shadowing terms from the MN $m, \forall m\in\mathcal{M}$ to the untrusted users $k \in\mathcal{K}$  is given by $\mathbb{E}\{A_{mk}A_{jk'}\} =
 4^22^{-\delta_{kk'}/9\,\text{m}}$, if $j=m$, and $\mathbb{E}\{A_{mk}A_{jk'}\} = 0$,  if $j \neq m$, $\forall j\in\mathcal{M},$
where $\delta_{kk'}$ is the  distance between UTs $k$ and $k'$.

\begin{figure}[t]
	\centering
	\vspace{-1.0em}
	\includegraphics[width=0.40\textwidth]{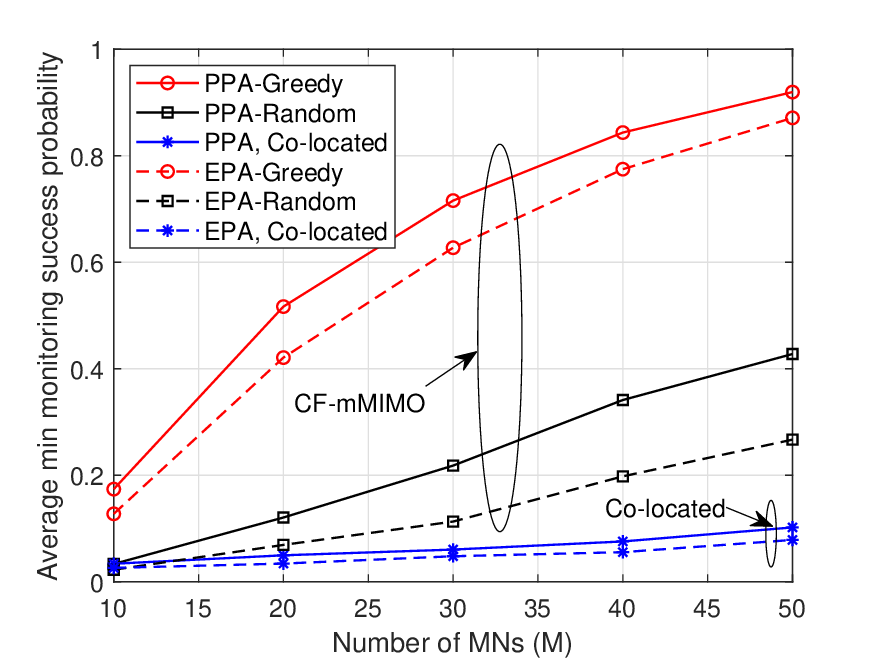}
	\vspace{-0.2em}
	\caption{Average minimum monitoring success probability  versus the number of MNs, $M$, where $K=10$.}
	\vspace{-0.5em}
	\label{fig:Fig1}
\end{figure}

Figure~\ref{fig:Fig1} illustrates the minimum monitoring success probability  achieved by the  CF-mMIMO surveillance system for different number of MNs $M$, where $K=10$. Our results verify the advantage of our proposed power allocation (PPA) and greedy mode assignment over equal power allocation (EPA)  and random mode assignment, respectively.  More specifically, PPA provides performance gains of up to $40\%$ and $100\%$ with respect to EPA for the system relying on greedy  mode assignment and random mode assignment, respectively. In addition, the remarkable performance gap between the greedy and random mode assignment verifies the importance of an adequate mode selection in terms of monitoring performance in the  CF-mMIMO surveillance system. In Fig.~\ref{fig:Fig1}, the  monitoring success probability of the CF-mMIMO and co-located mMIMO is also compared.  It can be observed that the CF-mMIMO  surveillance system  significantly outperforms its co-located mMIMO counterpart. For example, for $M=30$, the CF-mMIMO system provides around $11$-fold improvement in the minimum monitoring success performance  over the co-located system. The reason is twofold: primarily due to the ability of the former to surround each UT and each UR  by MNs in observing and jamming mode, respectively; secondarily, in contrast to our CF-mMIMO with distributed MNs, a co-located mMIMO  suffers from excessive self-interference due to the short distance between the transmit and receive antennas located at a single large MN.

\begin{figure}[t]
	\centering
	\vspace{-1.0em}
	\includegraphics[width=0.40\textwidth]{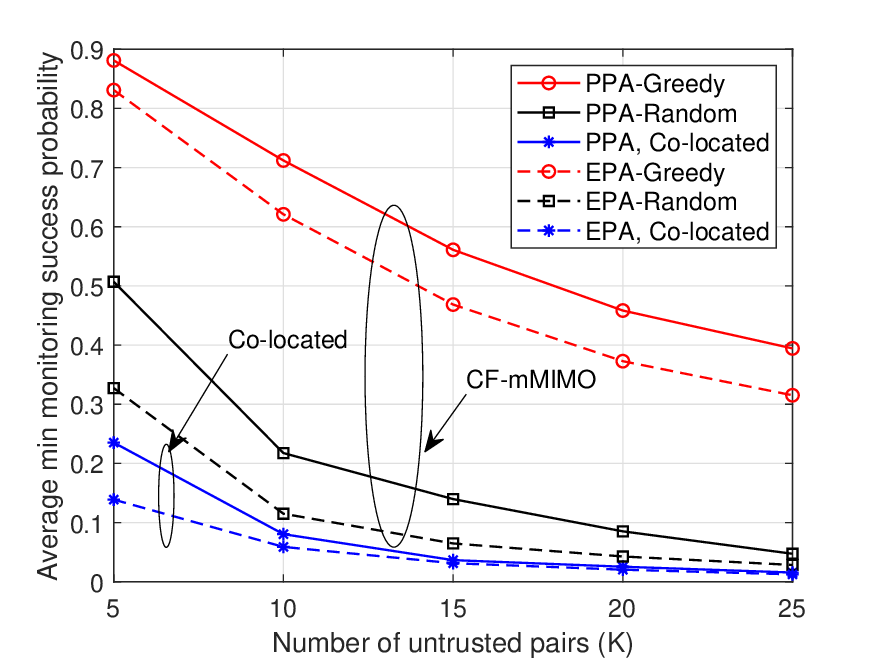}
	\vspace{-0.2em}
	\caption{Average minimum monitoring success probability  versus the number of untrusted pairs, $K$, where $M=30$.}
	\vspace{0em}
	\label{fig:Fi2}
\end{figure}

Next, in Fig.~\ref{fig:Fi2}, we investigate the impact of the number of untrusted pairs on the monitoring success probability performance. We observe that upon increasing $K$, the monitoring performance of all schemes deteriorates. Interestingly,  the relative performance gap between our CF-mMIMO system using the PPA and greedy mode assignment  and other schemes considerably increases with $K$. This result shows that the CF-mMIMO surveillance system along employing PPA and suitable mode assignment  guarantees  excellent monitoring performance in  practical multiple-untrusted-pair  scenarios.
\vspace{-0.4em}
\section{Conclusion}
We  designed CF-mMIMO surveillance systems for monitoring  multiple untrusted pairs and analyzed the  performance. We proposed a long-term-based greedy algorithm for designing the MN mode  assignment and a  power control technique for the MNs that are in jamming mode in order to maximize the minimum monitoring success probability  across all untrusted pairs  under a realistic transmit power constraint. We showed that our proposed schemes provide significant  monitoring gains over the random mode assignment and equal-power based schemes. Our results also confirmed that our CF-mMIMO surveillance system  significantly outperforms conventional co-located surveillance systems in terms of monitoring, even for relatively small number of MNs.
\vspace{-1.0em}
\appendices
\section{Proof of Theorem~\ref{Theorem:SE:SUS}}~\label{ProofTheorem:SE:SUS}
To derive the closed-form expression for  the effective SINR  of the untrusted link, we have to calculate $\xi_k(\qa,\boldsymbol{\theta})=\mathbb{E}\Big\{\big|\tilde w_{k}^{\UR}\big|^2\Big\}$. Denote   the jamming channel
estimation error by $\boldsymbol{\varepsilon}_{mk}^{\mathrm{J}}=\gmkd-\hgmkd$. Therefore, we have
 \begin{align}\label{eq:SINRD1}
&\xi_k(\qa,\boldsymbol{\theta})=\mathbb{E}\Big\{\big|\tilde w_{k}^{\UR}\big|^2\Big\}=
 \rho_{\UT}\sum_{\ell\in \mathcal{K}\setminus k} \betakldu+{\rho_\dl} \mathcal{I}+1, 
  \end{align}
  where
  \vspace{-0.4em}
 \begin{align*}
 \mathcal{I}\triangleq & \sum_{k'\in\mathcal{K} } \mathbb{E} \Big\{\Big|
\sum_{m \in \mathcal{M}}
 a_m\sqrt{\theta_{mk'}}
(\gmkd)^T\left(\hgmkpd\right)^*
\Big|^2\Big\}.
  \end{align*} 
To calculate $\mathcal{I}$, owing to the fact that the variance of a sum of
independent RVs is equal to the sum of the variances, we have
 \begin{align}\label{eq:SINRD2}
 &\mathcal{I}\stackrel{(a)}{=}
\sum_{k'\in\mathcal{K}\setminus k} \mathbb{E} \Big\{\Big| \!
 \sum_{m \in \mathcal{M}}\!
\! a_m\sqrt{\theta_{mk'}}\left(\gmkd\right)^T\left(\hgmkpd\right)^*\Big|^2\Big\}
\nonumber\\
 &\qquad+ \mathbb{E}\Big\{
\Big|\sum_{m \in \mathcal{M}}
a_m \sqrt{\theta_{mk}}
\left(\boldsymbol{\varepsilon}_{mk}^{\mathrm{J}}+\hgmkd\right)^T\left(\hgmkd\right)^*\Big|^2\Big\}\nonumber
 \\
&\!\!\!\stackrel{(b)}{=}\!\!\! \!
\sum_{k'\in\mathcal{K}\setminus k} 
 \sum_{m \in \mathcal{M}}\!\!
\!a_m {\theta_{mk'}}\mathbb{E}\! \Big\{\!\left(\gmkd\right)^{\!T}\!\!\mathbb{E} \Big\{\!\left(\hgmkpd\right)^*\!\!\left(\hgmkpd\right)^{\!T}\!\!\Big\}\left(\gmkd\right)^*\!\!\Big\}\nonumber
 \\
&\qquad+\sum_{m \in \mathcal{M}}a_m\theta_{mk}\big(\mathbb{E} \big\{\big\Vert\hgmkd\big\Vert^4\big\}+\mathbb{E} \big\{\big|{(\boldsymbol{\varepsilon}}_{mk}^{\mathrm{J}})^T(\hgmkd)^*\big|^2\!\big\}\!\big)\nonumber\\
&~~+\sum_{m \in \mathcal{M}}\sum_{n \in \mathcal{M}\setminus m}a_m a_n\sqrt{\theta_{mk}\theta_{nk}} \mathbb{E} \big\{\Vert\hgmkd\Vert^2\!\big\}
\mathbb{E} \big\{\Vert\hgnkd\Vert^2\!\big\}\nonumber
 \\
&\!\stackrel{(c)}{=}\! 
N
\!\! \sum_{k'\in\mathcal{K}}\!\sum_{m \in \mathcal{M}}\!\!\!\!
 a_m\theta_{mk'}\betamkd\gamdmkp  
  \!\!+\!\!N^2\big(\!\!\sum_{m \in \mathcal{M}}\!\!\!a_m \sqrt{\theta_{mk}} \gamdmk\! \big)^{\!2}\!\!,
  \end{align}

where $(a)$  follows from the fact that $\hgmkpd$ has zero mean and it is independent of $\gmkd$ and $\hgmkd$, $(b)$ follows from the fact that $\boldsymbol{\varepsilon}_{mk}^{\mathrm{J}}$   is independent of $\hgmkpd$ and it is a zero-mean RV
 and $(c)$ follows from  the fact that $\mathbb{E} \big\{\Vert\hgmkd\Vert^4\big\}=\Ntx(\Ntx+1)(\gamdmk)^2$.
Substituting~\eqref{eq:SINRD2} into~\eqref{eq:SINRD1} completes the proof.
\vspace{-0.4em}
\section{Proof of Theorem~\ref{Theorem:SE:CPU}}~\label{ProofTheorem:SE:CPU}
By substituting~\eqref{eq:ymul} into~\eqref{eq:rul}  and then employing the use-and-then-forget capacity-bounding methodology~\cite{Hien:cellfree}, the  SINR observed for the untrusted link $k$ can be written as
\vspace{0.3em}
\begin{equation}\label{eq:SINRST}
\SINR_{k}^\ul = \frac{{|{\text{DS}}_k ^\ul{|^2}}}{{{\text{BU}}_k ^\ul + {\text{UI}}_k ^\ul + {\text{MI}}_k ^\ul + {\text{AN}}_k ^\ul}},
\end{equation}
where  $\text{DS}_k ^\ul, \text{BU}_k ^\ul, \text{UI}_k ^\ul, \text{MI}_k ^\ul$,  and $\text{AN}_k ^\ul$  are the desired
signal, beamforming gain uncertainty, inter-untrusted user interference,
inter-MN interference, and additive noise, respectively, which can be written as
\begin{align*}
& \text{DS}_k ^\ul \triangleq \sqrt {\rho _\UT} \mathbb{E}\Big\{ \sum\limits_{m \in {\mathcal{M}}} (1-a_m)  {(\hgmlu)^\dag}\gmlu\Big\},
\\
&  \text{BU}_k ^\ul \triangleq \mathbb{E}\Big\{ \Big|\sqrt {{\rho _\UT}} \Big(\sum\limits_{m \in {\mathcal{M}}} (1-a_m)  {(\hgmlu)^\dag}\gmlu
\\ 
&\qquad\quad- \mathbb{E}\Big\{ \sum\limits_{m \in {\mathcal{M}}}(1-a_m)  {(\hgmlu)^\dag}\gmlu\Big\} \Big){\Big|^2}\Big\} ,
\\ 
& {\text{UI}}_k ^\ul \triangleq {\rho _\UT}\sum\limits_{{\ell \in {\mathcal{K}}\backslash k }} {\mathbb{E}\Big\{ \Big|\sum\limits_{m \in {\mathcal{M}}} (1-a_m) {(\hgmlu)^\dag}{\mathbf{g}}_{m\ell}^\ul{\Big|^2}\Big\} ,}
\\
& {\text{MI}}_k ^{\ul} \triangleq {\rho _\dl}\sum\limits_{\ell \in {{\mathcal{K}}}} \mathbb{E} \Big\{ \Big|\sum\limits_{m \in {\mathcal{M}}} \sum\limits_{i \in {\mathcal{M}}} (1-a_m)a_i\sqrt{\theta _{i\ell}}  
 \nonumber \\
&\hspace{3em}\times {({\mathbf{\hat g}}_{mk}^\ul)^\dag}{\qF_{mi}}{({\mathbf{\hat g}}_{i\ell}^\dl)^{\ast}}{\Big|^2}\Big\},
\\ 
& {\text{AN}}_k ^\ul \triangleq \sum\limits_{m \in {\mathcal{M}}} (1-a_m) \mathbb{E} \big\{ |{(\hgmlu)^\dag}{|^2}\big\}.  
\end{align*}

Denote   the observing channel estimation error which is independent of $\hat \qg_{mk }^\ul$ and  zero-mean RV by $\boldsymbol{\varepsilon}_{mk}^\ul=\qg_{mk }^\ul- \hat \qg_{mk }^\ul$.
By exploiting the independence between the channel
estimation errors and the estimates, we have
\begin{align} \label{eq:DS}
 {\text{DS}}_k ^\ul &= \sqrt {{\rho _\UT}}  \sum\limits_{m \in {\mathcal{M}}} (1-a_m)  \mathbb{E}\left\{{(\hgmlu)^\dag}(\hgmlu + {\boldsymbol{\varepsilon}}_{mk }^\ul)\right\}\nonumber\\ 
& = \sqrt {{\rho _\UT}} \sum\limits_{m \in {\mathcal{M}}}  (1-a_m) \Nrx\gamuml.
\end{align}
Additionally,  ${\text{BU}}_k ^\ul$ can be written as
\begin{align} \label{eq:BI}
& \text{BU}_k ^\ul 
= \!{\rho_\UT}\sum\limits_{m \in {\mathcal{M}}} \!(1\!-\!a_m)  \mathbb{E}\Big\{ |{(\hgmlu)^\dag}\gmlu - \mathbb{E}\{ {(\hgmlu)^\dag}\gmlu\} {|^2}\Big\} \nonumber
\\ 
 &= \!\rho_\UT\!\!\!\sum\limits_{m \in {\mathcal{M}}}
 \!(1-a_m)  \Big(\!{ \mathbb{E}}\Big\{ \big|{(\hgmlu)^\dag}\boldsymbol{\varepsilon}_{mk}^\ul \!+\! \Vert\hgmlu{\Vert^2}{\big|^2}\!\Big\} \nonumber
\!-\! \Nrx^2(\gamuml)^2\!\Big)\nonumber
\\
& \stackrel{(a)}{=} \rho _\UT \sum\limits_{m \in {\mathcal{M}}} (1-a_m)  \big(\mathbb{E}\big\{ |{(\hgmlu)^\dag}\boldsymbol{\varepsilon}_{mk}^\ul{|^2}\big\}   \nonumber
\\
&\hspace{10em}+\mathbb{E}\big\{ \Vert\hgmlu{\Vert^4}\big\}- \Nrx^2(\gamuml)^2\big) \nonumber
\\ 
&\stackrel{(b)}{=} \rho _\UT \sum\limits_{m \in {\mathcal{M}}} (1-a_m)  \Nrx\beta _{mk }^\ul\gamuml, 
\end{align}
where we have exploited: in $(a)$ $\boldsymbol{\varepsilon}_{mk}^\ul$  is independent of $\hat \qg_{mk }^\ul$ and a zero-mean RV; in $(b)$ $\mathbb{E} \big\{\Vert\hgmlu\Vert^4\big\}=\Ntx(\Ntx+1)(\gamuml)^2$.

By exploiting the fact that $\hgmlu$ is independent of ${\mathbf{g}}_{m\ell}^\ul$ for $k \neq \ell$, while $\hgmlu$,  ${\qF_{mi}}$, and ${\mathbf{\hat g}}_{i\ell}^\dl$ are independent, we can formulate  $\text{UI}_k ^\ul$ and $\text{MI}_k ^\ul$, respectively, as
\vspace{0em}
\begin{align} \label{eq:SI}
 {\text{UI}}_k ^\ul&= {\rho _\UT}\sum\limits_{\ell \in {{\mathcal{K}}}\backslash k } {\sum\limits_{m \in {\mathcal{M}}}  }(1-a_m) \Ntx\gamuml\beta _{m\ell}^\ul,\\ 
{\text{MI}}_k ^\ul  &= {\rho _\dl}\sum\limits_{\ell \in {\mathcal{K}}} \sum\limits_{m \in {\mathcal{M}}} \sum\limits_{i \in {\mathcal{M}}} (1-a_m) a_{i}\theta_{i\ell} 
\nonumber
\\
&\hspace{5em}\times
\mathbb{E}   \{ |{({\mathbf{\hat g}}_{mk }^\ul)^\dag}{{\mathbf{F}}_{mi}}{({\mathbf{\hat g}}_{i\ell}^\dl)^{\ast}}{|^2}\} \nonumber
\\
&~= {\rho _\dl}\sum\limits_{\ell \in {\mathcal{K}}} {\sum\limits_{m \in {\mathcal{M}}} {\sum\limits_{i \in {\mathcal{M}}}  } } \Ntx^2		(1-a_m) a_{i}\theta_{i\ell}   \gamuml \beta_{mi} \gamma_{i\ell}^{\dl}\label{eq:RI}.
\end{align}
 The substitution of~\eqref{eq:DS},~\eqref{eq:BI},~\eqref{eq:SI}, and~\eqref{eq:RI} into~\eqref{eq:SINRST} and inserting  $\text{AN}_k ^\ul\!=\! \Nrx\!\!\!\!\sum\limits_{m \in {\mathcal{M}}}(1-a_m) \gamuml$, yields~\eqref{eq:SINRMA}.

\vspace{0em}
\bibliographystyle{IEEEtran}


\end{document}